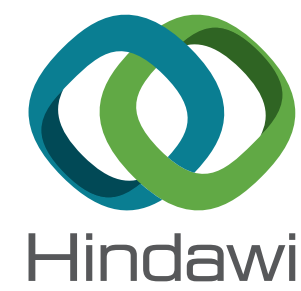

*Research Article*

# Lolisa: Formal Syntax and Semantics for a Subset of the Solidity Programming Language in Mathematical Tool Coq

**Zheng Yang and Hang Lei**

*School of Information and Software Engineering, University of Electronic Science and Technology of China, No.4 Section 2 North Jianshe Road, Chengdu 610054, China*

Correspondence should be addressed to Zheng Yang; zhengyang@std.uestc.edu.cn

The security of blockchain smart contracts is one of the most emerging issues of the greatest interest for researchers. This article presents an intermediate specification language for the formal verification of Ethereum-based smart contract in Coq, denoted as Lolisa. The formal syntax and semantics of Lolisa contain a large subset of the Solidity programming language developed for the Ethereum blockchain platform. To enhance type safety, the formal syntax of Lolisa adopts a stronger static type system than Solidity. In addition, Lolisa includes a large subset of Solidity syntax components as well as general-purpose programming language features. Therefore, Solidity programs can be directly translated into Lolisa with line-by-line correspondence. Lolisa is inherently generalizable and can be extended to express other programming languages. Finally, the syntax and semantics of Lolisa have been encapsulated as an interpreter in mathematical tool Coq. Hence, smart contracts written in Lolisa can be symbolically executed and verified in Coq.

## 1. Introduction

The blockchain platform [1] is one of the emerging technologies developed to address a wide range of disparate problems, such as those associated with cryptocurrency [2] and distributed storage [3]. Presently, this technology has gained interest from the finance sector [4]. Ethereum is one of the most widely adopted blockchain systems. One of the most important features of Ethereum is that it implements a very flexible general-purpose Turing-complete programming language denoted as Solidity [5]. This allows for the development of arbitrary applications and scripts that can be executed in a virtual runtime environment denoted as the Ethereum Virtual Machine (EVM) to conduct blockchain transactions automatically. These applications and scripts (i.e., programs) are collectively denoted as smart contracts, which have been widely used in many critical fields, such as the medical [6] and financial fields. The growing use of smart contracts has led to an increased scrutiny of their security. Smart contracts can include particular properties (i.e., bugs) making them susceptible to deliberate attacks that can result in direct economic loss. Some of the largest attacks on smart contracts are well known, such as the attack on decentralized autonomous organization (DAO) and Parity wallet [7] contracts. In fact, many classes of subtle bugs, ranging from transaction-ordering dependencies to mishandled exceptions, exist in smart contracts [8].

The present article capitalizes upon our past work by defining the formal syntax and operational semantics for a large subset of the Solidity version 0.4. This subset is denoted herein as Lolisa and has the following features.

*Consistency* Lolisa formalizes most of the types, operators, and mechanisms of Solidity according to Solidity documentation. As such, programs written in Solidity can be translated into Lolisa, and vice versa, with a line-by-line correspondence without rebuilding or abstracting, which are operations that can negatively impact consistency.

*Static Type System* The formal syntax in Lolisa is defined using generalized algebraic datatypes (GADTs) [9], which impart static type annotation to all the values and expressions of Lolisa. In this way, Lolisa has a stronger static type system than Solidity for checking the construction of programs.

*Executable and Provable* In contrast to similar efforts focused on building formal syntax and semantics for high-level programming languages, the formal semantics of Lolisa are defined based on the GERM framework in conjunction with EVI. Therefore, it is theoretically possible for ethereum-based smart contracts written in Lolisa to be symbolically executed and have their properties simultaneously verified automatically in higher-order logic theorem-proving assistants directly when conducted in conjunction with a formal interpreter developed based on GERM framework.

*Mechanized and Validated* The syntax and semantics of Lolisa are mechanized using the Coq proof assistant [10]. We also develop a formal verified interpreter in Coq to validate whether Lolisa satisfies the above *Executable and Provable* feature and the meta-properties of the semantics. The details regarding the implementation of our formal interpreter have been presented in another paper [11].

The remainder of this paper is structured as follows. Section 2 introduces related work regarding the programming language formalization. Section 3 introduces the overall structure of the specification language framework and provides predefinitions of Lolisa syntax and semantics. Section 4 elaborates on the formal abstract syntax of Lolisa and compares this with the formal abstract syntax of Solidity. Section 5 presents the formal dynamic semantics of Lolisa, including the program execution semantics and the formal standard library for the built-in data structures and functions of EVM. Section 6 describes the integration of the Lolisa programming language and its semantics within the formal verified interpreter FEther. Section 7 discusses the contributions and limitations of our current work. Finally, Section 8 presents the conclusions of our work.

## 2. Related Work

Software engineering techniques employing such static and dynamic analysis tools as Manticore [12] and Mythril [13] have not yet been proven to be effective at increasing the reliability of smart contracts.

KEVM [14] is a formal semantics for the EVM written using the K-framework, like the formalization conducted in Lem [15]. KEVM is executable, and therefore can run the validation test suite provided by the Ethereum foundation. The symbolic reasoning conducted for KEVM programs involves specifying properties in Reachability Logic and verifying them with a separate analysis tool. While these represent currently available mechanized formalizations of operational semantics, axiomatic semantics, and formal low-level programming verification tools for EVM and Solidity bytecode [16], they are not well-suited for high-level programming languages, such as Solidity. In response, the Ethereum community has placed open calls for formal verification proposals [17] as part of a concerted effort to develop formal verification strategies [18]. Fuzzing testing is an efficient and effective testing technique. Presently, numerous projects develop fuzzing in smart contracts to analyze vulnerabilities, such as ReGuard [19]. Securify [20] is a type of Ethereum-based smart contracts security analyzer based on static analysis. It verifies the behavior of target smart contracts based on the given security properties at the Ethereum virtual machine bytecode level. Securify provides a kind of domain-specific language which can write security properties according to the attack reports and the basic practices. MadMax [21] is a static program analysis framework that takes the Ethereum bytecode as analysis source code and automatically analyzes common vulnerabilities such as the integer and memory overflows vulnerabilities. Besides, it is the first tool that allows for loop specifications to be defined by a dynamic property. In this manner, this tool can avoid loop explosion during the verification process. Similarly to OYENTE, Ehtir [22] is also a type of rule-based static analyzer for the bytecode of Ethereum smart contracts. This tool can produce control flow graphs and includes the whole possible execution addresses. VeriSolid [23] is a formal verification framework which can be accessed through the web directly. Its foundational concept is FSolidM [24]. In brief, the VeriSolid presents a formal verification framework which provides an approach for semiautomatically developing the correct formal specifications of smart contracts. A new approach is presented in Abdellatif and Brousmiche [25] which can model the execution behaviors of target smart contracts based on a formal model checking language. This technique can be applied to verify the execution behavior and authority of target smart contracts by using model checking methods.

In other fields of computer science, a number of interesting studies have focused on developing mechanized formalizations of operational semantics for different high-level programming languages. The Park project [26] presents completely formalized denotational semantics and the corresponding syntax in the JavaScript language. The CompCert project [27] is another influential verification work for C and GCC that developed a formal semantics for a subset of C denoted as Clight. This work formed the basis for VST [28] and CompCertX [29]. In addition, a number of interesting formal verification studies have been conducted for operating systems based on the CompCert project. In addition, the operational semantics of JavaScipt also have been investigated [30], which is of particular importance to the present study because Solidity is a programming language like JavaScipt. However, few of the frameworks defined in these related works can be symbolically executed or analyzed in higher-order logic theorem-proving assistants directly.

## 3. Foundational Concepts

The overall architecture of Lolisa is shown in Figure 1. Table 1 summarizes the helper functions used in the dynamic semantic definitions. Table 2 lists the state functions used to calculate commonly needed values from the current state of the program. All of these state that functions will be encountered in the following discussion. Components of specific states will be denoted using the appropriate Greek letter subscripted by the state of interest. As shown in Table 2, the context of the formal memory space is denoted as $M$, where $\sigma$ is employed to denote a specific memory state; the context of the execution environment is represented as $\varepsilon$; and we assign $\Lambda$ to denote a set of memory addresses, where the meta-variable $\alpha$ is employed to represent an arbitrary address. Similarly, we define the function return address $\Lambda_{fun}$. In addition, struct is an important data structure in Lolisa. Therefore, we adopt $\Sigma$ to represent the Lolisa struct information context, and $\Theta$ is employed to represent the set of pointers of the struct types. Also, the following type of assignments may include variables, so our types will include references to variable-typing contexts, which we will denote as $\Gamma$, $\Gamma_1$, etc. Such contexts are finite mappings from variable names to types. Because programs may also contain references to the declared functions of a Solidity program, another mapping is needed from function identifiers to types. This mapping will be succinctly denoted as $\Phi$, $\Phi_1$, etc. Furthermore, we assign $\Omega$ as the native value set of the basic logic system. For brevity in the following discussion, we will assign $\mathcal{F}$ to represent the overall formal system combination of $\Sigma$, $\Gamma$, $\Theta$, $\Omega$, $\Phi$, and $\Lambda$. Due to limitation of length, the details of Lolisa's formalization have been presented in our online report (https://arxiv.org/abs/1803.09885).

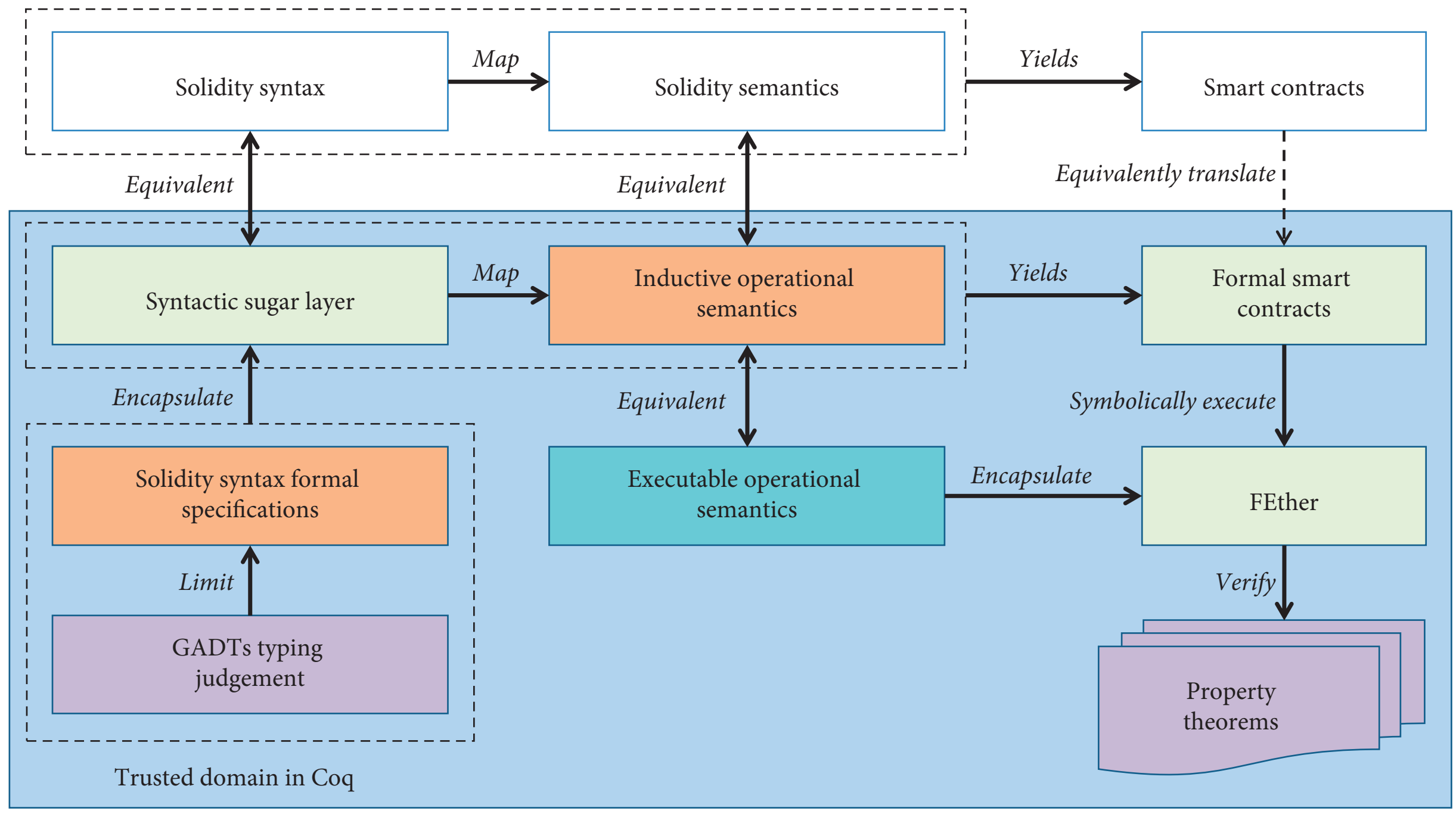

Figure 1: Overview of Lolisa's architecture.

Table 1: Helper functions.

| Symbol | Definition | Symbol | Definition |
| --- | --- | --- | --- |
| $map_{addr}$ | Searches the indexed address of a mapping type | $map_{get}$ | Obtains the value in a mapping type term |
| $eval_{bop}$ | Evaluates binary operation expressions | $eval_{uop}$ | Evaluates unary operation expressions |
| $mems_{find}$ | Searches the required struct member | $env_{check}$ | Validates the current environment |
| $set_{env}$ | Changes the current environment | $inherit_{chec}$ | Validates the inheritance information |
| $init_{var}$ | Initializes the variable address | $init_{re}$ | Initializes the function return address |

Table 2: State functions.

| Symbol | Definition | Symbol | Definition |
| --- | --- | --- | --- |
| $M$ | Memory space | $\varepsilon$ | Environment information |
| $\Lambda$ | Memory address set | $\Lambda_{fun}$ | Function return address |
| $\Sigma$ | Struct information | $\Theta$ | Struct pointer set |
| $\Gamma$ | Context structure information | $\Omega$ | Native value set |
| $\Phi$ | Function information | $\mathcal{F}$ | Overall formal system |

## 4. Formal Syntax of Lolisa

*4.1. Types.* The formal abstract syntax of Lolisa types is given in Figure 2. Supported types include arithmetic types (integers in various sizes and signedness), byte types, array types, mapping types, as well as function types and struct types. Although Solidity is a JavaScript-like language, it supports pointer reference. Therefore, Lolisa also includes pointer types (including pointers to functions) based on label address specification. Furthermore, these types of annotations and relevant components can be easily formalized by enumerating inductively in Coq or other higher-order logic theorem-proving assistants. Lolisa does not support any of the type qualifiers such as const, volatile, and restrict, and these qualifiers are simply erased during parsing.

| | |
|---|---|
| *Signedness*: | $signedness ::= Signed \mid Unsigned$ |
| *Integer sizes*: | $intsize ::= I8 \mid I16 \mid I32 \mid I64 \mid I128 \mid I256$ |
| *Byte sizes*: | $bytesize ::= B4 \mid B8 \mid B16 \mid B32$ |
| *Array Index of map* | $id_{map_{array}} ::= iAconst_id(n) \mid iAvar_id(\alpha)$ |
| | $\mid iAstr_id(\alpha, mems) \mid iArray_id(\alpha, id_{map_{array}})$ |
| *Normal* $type_{map}$: | $\tau_{nf_{map}} ::= Iundef \mid Iint(signedness, intsize)$ |
| | $\mid Ibool \mid Istring \mid Ifloat \mid Ibytes(bytesize) \mid Istruct(\alpha)$ |
| | $\mid Ivid(o\alpha) \mid Ipid(o\alpha) \mid Ifid(o\alpha) \mid Icid(o\alpha)$ |
| *Map Types Sig*: | $\tau_{map} ::= Iarray(id_{map_{array}}, \tau_{map}) \mid \tau_{nf_{map}}$ |
| | $Iaddress ::= Istruct(_0xaddress) \mid IInt ::= Iint\ Signed\ I64.$ |
| *Array Index* | $id_{array} ::= Aconst_id(n) \mid Avar_id(\alpha)$ |
| | $\mid Astr_id(\alpha, mems) \mid Amap_id(\alpha, id_{map_{array}}) \mid Array_id(\alpha, id_{array})$ |
| *Normal type*: | $\tau_{nf} ::= Tundef \mid Tint(signedness, intsize)$ |
| | $\mid Tbool \mid Tstring \mid Tfloat \mid Tbytes(bytesize) \mid Tstt$ |
| | $\mid Tmodi \mid Tstruct(\alpha) \mid Tvid(o\alpha) \mid Tpid(o\alpha) \mid Tfid(o\alpha) \mid Tcid(o\alpha)$ |
| *Types Sig*: | $\tau ::= Tmap(\tau_{map}, \tau) \mid Tarray(id_{array}, \tau) \mid \tau_{nf}$ |
| | $Taddress ::= Tstruct(_0xaddress) \mid TInt ::= Tint\ Signed\ I64$ |
| | $Tuint ::= Tint\ Unsigned\ I64$ |

Figure 2: Abstract syntax of Lolisa types.

The types fill two roles in Lolisa. Firstly, they serve as type declarations of identifiers in statements and, secondly, they serve as signatures to specify the GADTs-style constructor of values and expressions for transmitting type information, which will be explained in the following sections. In Coq formalization, the term $\tau$ is declared as type according to rule 1, as follows:

$$\tau\text{: type.} \tag{1}$$

Note that many types are defined in Figure 2 as parameterized types recursively. In this way, a specific type is dependent on the specified parameters and can abstract and express many different Solidity types.

One of the most important data types of Solidity is mapping types. In Solidity documentation [4], mapping types are declared as mapping (KeyType⇒ValueType). Here, _KeyType can be nearly any type except for a mapping, a dynamically sized array, a contract, and a struct. As shown in Figure 2, _KeyType is defined as Tmap ($\tau_{\text{map}}$, $\tau$), where $\tau_{\text{map}}$ represents the _KeyType and $\tau$ represents the _ValueType. The best way to keep the terms in Lolisa well-typed and to ensure type safety is to maintain type isolation rather than adding corollary conditions. Therefore, we define a coordinate type $\text{type}_{\text{map}}$ for _KeyType employed in mapping. In particular, the address types in Lolisa are treated as a special struct type, so that _KeyType is allowed to be a struct type in Lolisa. In Coq formalization, $\text{type}_{\text{map}}$ shares the same constructor with that of type except for Tmap, and a term with type $\text{type}_{\text{map}}$ is recorded as $\tau_{\text{map}}$ according to rule 2, as follows:.

$$\tau_{\text{map}}\text{: type}_{\text{map}}. \tag{2}$$

In Solidity, array types, which are defined according to an array index $\text{id}_{\text{array}}$ as Tarray ($\text{id}_{\text{array}}$, $\tau$) in Coq, can be classified as fixed-size arrays and dynamic-size arrays. For fixed-size arrays, the size and index number are allowed to be declared by different data structures including constants, variables, struct, mapping, and field access values. These are respectively formalized as Array Index in Figure 2. Because the size of array types in Solidity can be dynamic, the dynamic-size array type in Lolisa is treated as a special mapping type of $\tau_{\text{map}}$ (Iint Signed I64).

As shown in Figure 3, (n)-dimensional mapping types, as well as array types, are widely defined in smart contracts. Due to the recursive inductive definition, Lolisa can express

```
1contract MappingExample{
2    mapping(address => mapping(char => uint)) public balances;
3        function update(uint amount) returns (address addr){
4        ...;
5        return msg.sender;
6    }
7}
```

FIGURE 3: A simple example of mapping types in Solidity.

n-dimensional array types and n-dimensional mapping types easily, which is illustrated below by rules 3 and 4, respectively:

$$\mathrm{Tarray}\left[\mathrm{id}_0\mathrm{Tarray}\left[\mathrm{id}_1\mathrm{Tarray}\left[\ldots\left[\mathrm{Tarray}\,\mathrm{id}_n\tau_{\mathrm{final}}\right]\right]\right]\right], \tag{3}$$

$$\mathrm{mapping}\left[\tau_{\mathrm{map}_0}\Rightarrow\mathrm{mapping}\left[\tau_{\mathrm{map}_0}\Rightarrow\mathrm{mapping}\left[\ldots\mathrm{mapping}\left[\tau_{\mathrm{map}_n}\Rightarrow\tau_{\mathrm{final}}\right]\right]\right]\right]. \tag{4}$$

We classify $\tau$ and $\tau_{\mathrm{map}}$ into normal form types and nonnormal form types. The normal form types refer to types whose typing rules disallow recursive definition, whereas recursive definition is allowed for nonnormal form types. For example, the normal form of $\mathrm{Tarray}\,(\mathrm{id}_{\mathrm{array}}, \mathrm{Tbool})$ should be Tbool. In Figure 2, the normal types are defined separately as *Normal type*.

*4.2. Expressions.* Having formally specified all the possible forms of values that may be declared and manipulated in Solidity programs, we now discuss the expressions used in programs to encapsulate values. As introduced in Section 4.1, all expressions and their subexpressions are defined with GADTs, which are annotated by two types of signatures according to rule 5, as follows:

$$\mathrm{expr}\colon\ \tau_0 \longrightarrow \tau_1 \longrightarrow \mathrm{Type}. \tag{5}$$

Here, $\tau_0$ refers to the current expression type and $\tau_1$ refers to the normal form type after evaluation. For instance, we would define the type of an integer variable expression $e$ as $\mathrm{expr}_{\mathrm{Tvid}_{o\alpha}}$. In this way, the formal syntax of expressions becomes clearer and abstract, and allows the type safety of Lolisa expressions to be maintained strictly. In addition, employing the combination of the two types of annotations facilitates the definition of a very large number of different expressions based on equivalent constructors. Of course, the use of $\tau_0$ and $\tau_1$ may be subject to different limitations depending on the situation.

Constant expressions are used to denote the native values of the basic formal system, which are transformed from the respective Lolisa values. Therefore, $\tau_0$ and $\tau_1$ should satisfy rule 6 given below:

$$\begin{aligned}\tau_0 &= \tau_1 \Lambda \tau_0,\\ \tau_1 &\in \tau_{nf}.\end{aligned} \tag{6}$$

To satisfy the limitation TYPE-FORM, the array types and mapping types should be analyzed and simplified according to the type definitions given by Figure 2 into $\tau_{\mathrm{final}} \in \tau_{\mathrm{nf}}$, which can be formulated as $\Sigma\Theta\leftarrow\tau \longrightarrow \tau' \longrightarrow \cdots \longrightarrow \tau^{\mathrm{n}}\Lambda\tau^{\mathrm{n}} \in \tau_{\mathrm{nf}}$. We denote this process as $\Downarrow_\tau$.

In addition, as mentioned previously, the type information of the value level is successfully transmitted into a constant expression. For example, a value $v$ has type $\mathrm{val}\,\tau_1$, and the constant expression Econst has type $\forall\,(\tau : \mathrm{type}),\ \mathrm{val}\,\tau \longrightarrow \mathrm{expr} \Downarrow_\tau \Downarrow_\tau$. Therefore, $\tau$ in $\mathrm{Econst}\,(\mathrm{v})$ is determined by $\tau_1$. For example, $\mathrm{Econst}\,(\mathrm{Vbool}\,(\mathrm{b}))$ has type $\mathrm{expr}\,\mathrm{Tbool}\,\mathrm{Tbool}$, where $\tau$ is specified by the Tbool of $\mathrm{Vbool}\,(\mathrm{b})$. The type information of the expression level can also be transmitted to the statement level in the same way, which will be described specifically in the next section.

For operator expressions, Lolisa supports nearly all binary and unary operators and we adopt $\mathrm{op}_{\mathrm{class}}$ (operator) to simplify the formal abstract syntax. In Coq formalization, binary and unary operators are abstracted as an inductive type op that is also defined by GADTs, and specific operators serve as their constructors. In this way, operator expressions are made more clear and concise, and can be extended more easily than when employing a weaker static-type system. The binary and unary operators are annotated by two type signatures, as respectively given in rule 7, as follows:

$$\mathrm{op}\colon\ \tau_0 \longrightarrow \tau_1 \longrightarrow \mathrm{Type}. \tag{7}$$

*4.3. Statements.* Figure 4 defines the syntax of Lolisa statements. Here, nearly all the structured control statements of Solidity (i.e., conditional statements, loops, structure declarations, modifier definitions, contracts, returns, multivalue returns, and function calls) are supported, but Lolisa does not support unstructured statements such as goto and unstructured switches like the infamous "Duff's device". Besides, anonymous functions are forbidden in Lolisa because all functions must have a binding identifier to ensure that they are well formed. As previously discussed, the assignment $e_1 = e_2$ of a right-value (r-value) $e_2$ to a left-value (l-value) $e_1$, and modifier declarations, as well as function calls and structure declarations are treated as

$$
\begin{aligned}
& & & strt_{mem} ::= \mathit{str_nil}(\mathit{addr}) \mid \mathit{str_mem}(\tau, \mathit{name}, \mathit{strmem}) \\
& & & stt_{nf} ::= \mathit{Var}(\mathit{oacc}, \varepsilon_{Tvar}) \mid \mathit{Struct}(a, strt_{mem}) \mid \mathit{Assignv}(\varepsilon_{addr}, \varepsilon) \\
& & & \mid \mathit{Return}(\varepsilon) \mid \mathit{Returns}(\varepsilon_s) \mid \mathit{Throw} \mid \mathit{Snil} \mid \mathit{Fstop} \\
& \mathit{Statement:} & s ::= & \mathit{Contract}(\varepsilon_{Tpar}) \mid \mathit{Modifier}(\varepsilon_{Tpar}) \\
& & & \mid \mathit{Fun}(\mathit{oacc}, \mathit{oflag}, \mathit{opars}\ \varepsilon_{Tfun}, \mathit{modis}, s) \\
& & & \mid \mathit{Funs}(\mathit{oacc}, \mathit{oflag}, \mathit{opars}, \varepsilon_{Tfun}, \mathit{ts}, \mathit{modis}, s) \\
& & & \mid Loop_{while}(\varepsilon_{Tbool}, s) \mid Loop_{for}(s_1, \varepsilon_{Tbool}, s_2, s_3) \\
& & & \mid Fun_{call}(\tau, \tau', \mathit{fitype}, \mathit{fipar}) \mid \mathit{Seq}(s, s') \mid \mathit{If}(\varepsilon_{Tbool}, s, s') \mid stt_{nf}
\end{aligned}
$$

FIGURE 4: Abstract syntax of Lolisa statements.

statements. In addition, statements are also classified according to normal form and nonnormal form categories, where the normal form statement, given as $\text{stt}_{\text{nf}}$, represents a statement that halts after being evaluated. Actually, while Solidity is a Turing-complete language, smart contract programs written in Solidity have no existing halting problems because program execution is limited by gas, which we have defined in $\varepsilon$ for Lolisa.

As defined in Figure 4, we still inductively classify statement definitions into a normal form $\text{stt}_{\text{nf}}$, whose typing assignments must be conducted without recursive definition, and non-normal form statements. The normal form statements of Lolisa are defined as $\text{stt}_{\text{nf}}$. The remaining statements are nonnormal form statements.

*4.4. Macro Definition of Formal Abstract Syntax.* The Lolisa formal syntax is too complex to be adopted by general users. Lolisa syntax includes the same components as those employed in Solidity; however, it has stricter formal typing rules. Therefore, Lolisa syntax must include some additional components not supported in Solidity, such as type annotations and a monad-type option. Moreover, Lolisa syntax is formally defined in Coq formalization as inductive predicates. Thus, a Lolisa code looks much more complicated than the corresponding Solidity code, even though both the codes demonstrate line-by-line correspondence. An example of this difficulty is illustrated in the code segments shown in Figures 5 and 6. The formal Lolisa version of the conditional statement in the *pledge* function in Figure 6 is much more complicated than that in the original Solidity version in Figure 5.

The degree of complexity poses a challenge for general users to write Lolisa codes manually and develop a translator between Lolisa and Solidity or another language. This is a common issue in nearly all similar higher-level language formalization studies.

Fortunately, Coq and other higher-order theorem-proving assistants provide a special macro-mechanism. In Coq, this mechanism is referred to as the notation mechanism. Here, a notation is a symbolic abbreviation denoting a term or term pattern automatically parsed by Coq. For example, the symbols in Lolisa can be encapsulated as shown in Figure 7.

The new formal version of this example yields the notation in Figure 8, which demonstrates that the notation is nearly equivalent to the original Solidity syntax.

Through this mechanism, we can hide the fixed formal syntax components used in verification and thereby provide users with a simpler syntax. Moreover, this mechanism makes the equivalence between real-world languages and Lolisa far more intuitive and user friendly. In addition, this mechanism improves verification automation. Similar to converting Figures 5–8, we develop a translator, constructed by a lexical analyzer and a parser, to automatically convert the Solidity program to the macro definitions of the Lolisa abstract syntax tree. The translation process is given in Figure 9. The textual scripts of Ethereum smart contracts will be analyzed by the lexical analyzer of translator, which will generate the Solidity token stream. According to the syntactic sugar of Lolisa, the lexical analyzer will generate the respective Lolisa token stream. Next, the parser will take the Solidity token stream as parameters and generate the parse tree of smart contracts. Finally, the tokens of the parse tree will be replaced by the Lolisa token stream, and then the parser will rebuild the Lolisa parse tree and output the respective formal smart contracts rewritten by Lolisa. In this manner, the translation process can be guaranteed to be completed mechanically.

## 5. Formal Semantics

*5.1. Evaluation of Expressions.* The semantics of expression evaluation are the rules governing the evaluation of Lolisa expressions into the memory address values of the GERM framework, and this process includes two parts: the l-value position evaluation and the r-value position evaluation. In contrast, modifier expressions are a special case that cannot be evaluated according to these expression evaluation semantics, but their evaluation is conducted according to rule 8:

$$
\vdash_{\Downarrow} \langle \sigma, \text{env}, \text{fenv}, E\text{modifier}(\,*\,) \rangle \Rightarrow \langle \sigma, \text{env}, \text{fenv}, \text{Error} \rangle. \tag{8}
$$

Here, $\Downarrow_{\text{e}}$ represents the process of evaluating a modifier expression both in the l-value position and the r-value position. And the example semantics are summarized in Figure 10.

```
1function pledge(bytes32 _message) {
2    if (msg.value ==0 || complete || refunded) throw;
3    ...;
4}
```

Figure 5: Conditional statement in Solidity.

```
1Definition fun_pledge :=
2(Fun (Some public) (Efun (Some pledge ) Tundef )
3  ( pcons (Epar (Some _message) Tbyte32) pnil))
4  ( nil )
5  (
6    (If ((Ebop forbOfBool
7           (Ebop forbOfBool
8            (Ebop feqbOfInt
9              (Econst (Vfield Tuint (Fstruct _0xmsg msg) (@cons str_name (Nvar values) (@nil str_name)) None))
10             (Econst (Vint (INT I64 Signed 0))))
11            (Evar (Some complete) Tbool))
12          (Evar (Some refunded) Tbool)
13          )
14        )
15        (Throw;;;) (Snil;;;)
16    );;;
17    ...
18);;;
19.
```

Figure 6: Formal version of the conditional statement shown in Figure 5 in Lolisa.

```
1Notation "s '~>' ss" := (@cons str_name (Nvar s) ss)
2    (at level 59, right associativity).
3Notation " '\\' " := (@nil str_name)
4    (at level 59, right associativity).
5...
6Notation " e1 '(||)' e2 " := (Ebop forbOfBool e1 e2)
7    (at level 58, right associativity).
8Notation " e1 '(<)' e2 " := (Ebop fltbOfInt e1 e2)
9    (at level 52, right associativity).
10Notation " e1 '(==)' e2 " := (Ebop feqbOfInt e1 e2)
11    (at level 53, right associativity).
12...
```

Figure 7: Macro definitions of Lolisa formal abstract syntax tree.

```
1Definition fun_pledge :=
2(Fun (Some public) (Efun (Some pledge ) Tundef )
3  ( pcons (Epar (Some _message) Tbyte32) pnil))
4  ( nil )
5  (
6    (If ((SCField Tuint Msg (values ~> \\) None) (==) (SCInt 0) (||) (SVBool complete) (||) (SVBool refunded))
7        (Throw;;;) (Snil;;;)
8    );;;
9  ...
10);;;
11.
```

Figure 8: Simplified formal version of Figure 5 using syntactic abbreviations.

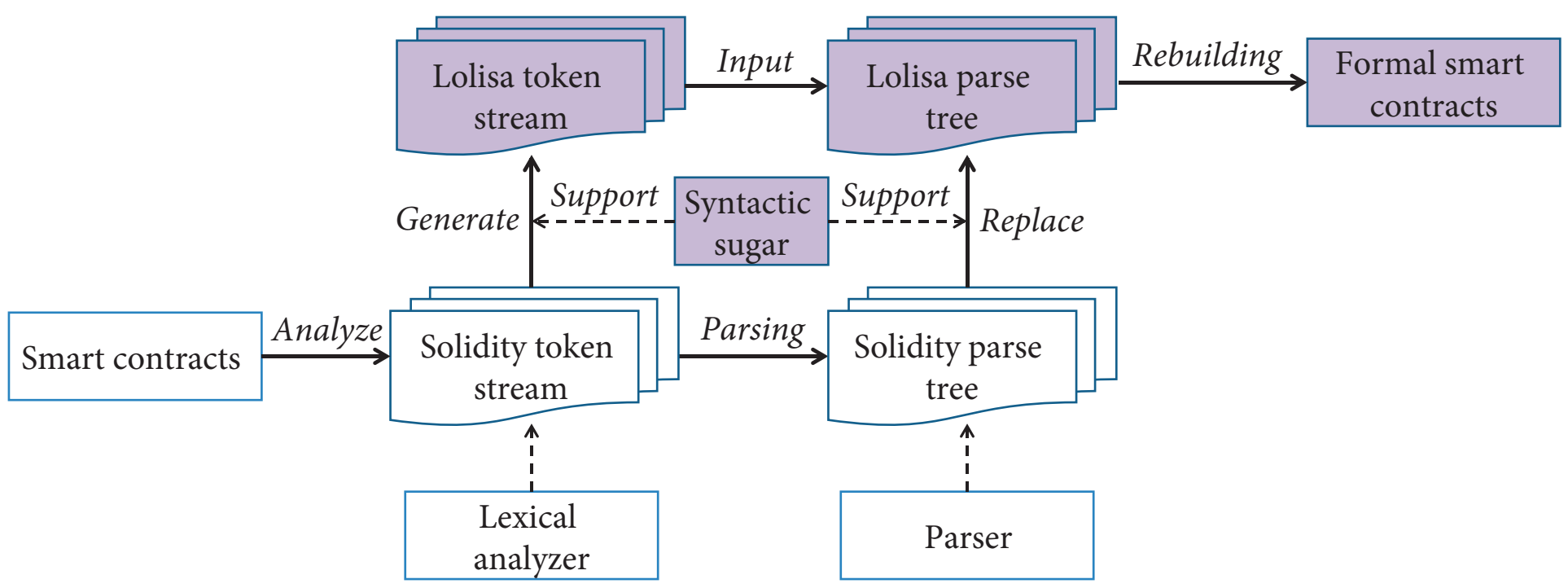

Figure 9: Translation process from smart contracts to its formal version.

$$\dfrac{\begin{array}{c}\mathcal{E} \vdash env, fenv \qquad M \vdash \sigma, b_{infor} \\ \mathcal{F}, b_{infor} \vdash v = Varray\left(name\left[id_{array}\right]\tau\right) \\ \mathcal{E}, M, \mathcal{F} \vdash \mathcal{A}\left]\sigma, name, \Downarrow_{id_{array}}, \tau, address_{offset}\right] \xrightarrow{yields} Some\ addr\end{array}}{\mathcal{E}, M, \mathcal{F} \vdash \langle \sigma, env, fenv, Econst(v)\rangle \Rightarrow \langle \sigma, env, fenv, Some\ addr\rangle}\text{(EVAL-LEXP-ARR)}$$

$$\dfrac{\begin{array}{c}\mathcal{E} \vdash env, fenv \qquad M \vdash \sigma, b_{infor} \\ \mathcal{F}, b_{infor} \vdash v = Vmap\left(name\left[id_{map}\right]\left(\tau_{map} \Rightarrow \tau\right) snd\right) \\ \mathcal{E}, M, \mathcal{F} \vdash \mathcal{A}\left]\sigma, name, \Downarrow_{id_{map}}, \tau, \tau_{map}, snd\right] \xrightarrow{yields} Some\ addr\end{array}}{\mathcal{E}, M, \mathcal{F} \vdash \langle \sigma, env, fenv, Econst(v)\rangle \Rightarrow \langle \sigma, env, fenv, Some\ addr\rangle}\text{(EVAL-EXP-MAP)}$$

$$\dfrac{\begin{array}{c}M \vdash \sigma, b_{infor} \qquad \mathcal{F}, b_{infor} \vdash v \\ \mathcal{E}, M, \mathcal{F} \vdash \sigma \Downarrow_{v} \Rightarrow (\sigma, Some\ m_v)\end{array}}{\mathcal{E}, M, \mathcal{F} \vdash \langle \sigma, env, fenv, Econst(v)\rangle \Rightarrow \langle \sigma, env, fenv, Some\ m_v\rangle}\text{(EVAL-REXP-CONS)}$$

$$\dfrac{\begin{array}{c}\mathcal{E} \vdash env, fenv \qquad M \vdash \sigma, b_{infor} \qquad \mathcal{F}, b_{infor} \vdash mems \stackrel{\text{def}}{=} \{v_0; v_1; \ldots; v_n\} \\ \mathcal{E}, M, \mathcal{F} \vdash \forall i \in \mathbb{N}. 0 \leq i \leq n,\ \sigma_i \Downarrow_{v_i} \Rightarrow \left(\sigma_i{}', out_i\right) \\ \forall i \in \mathbb{N}. 0 \leq i \leq n, out_i \neq Error \supset mems = Some\left[out_0.(m_{v_0}); out_1.(m_{v_1}); \ldots; out_n.(m_{v_n})\right]\end{array}}{\mathcal{E}, M, \mathcal{F} \vdash \langle \sigma, env, fenv, Estruct(head\ \{v_0; v_1; \ldots; v_n\})\rangle \Rightarrow \langle \sigma, env, fenv, Some\left(Str\left(head, mems, env, b_{infor}\right)\right)\rangle}\text{(EVAL-REXP-STR)}$$

$$\dfrac{\begin{array}{c}\mathcal{E} \vdash env, fenv \qquad M \vdash \sigma \\ \mathcal{E}, M, \mathcal{F} \vdash \sigma \Downarrow_{e_0}^{r}\left(\sigma, Some\ m_{v_0}\right) \wedge \sigma \Downarrow_{1}^{r}\left(\sigma, Some\ m_{v_1}\right) \qquad eval_{bop}(\sigma, op_2, m_{v_0}, m_{v_1}) \hookrightarrow re\end{array}}{\mathcal{E}, M, \mathcal{F} \vdash \langle \sigma, env, fenv, Ebop(op_2, e_0, e_1)\rangle \Rightarrow \langle \sigma, env, fenv, re \sqsubseteq \{Error, Some\ m_v'\}\rangle}\text{(EVAL-REXP-BOP)}$$

$$\dfrac{\begin{array}{c}\mathcal{E} \vdash env, fenv \qquad M \vdash \sigma \\ \sigma \Downarrow_{e_0}^{r} \Rightarrow Some\ m_{v_0} \qquad eval_{uop}(\sigma, op_1, m_{v_0}) \hookrightarrow re\end{array}}{\mathcal{E}, M, \mathcal{F} \vdash \langle \sigma, env, fenv, Euop(op_1, e_0)\rangle \Rightarrow out\ \langle \sigma, env, re \sqsubseteq \{Error, Some\ m_v'\}\rangle}\text{(EVAL-REXP-UOP)}$$

Figure 10: Formal operational semantics of Lolisa left and right expressions, including the array, mapping, constant, struct, and binary and unary operators.

*5.1.1. Evaluating Expressions in the L-Value Position.* In the following, we assign $\Downarrow_{e}^{l}$ to denote the evaluation of expressions in the l-value position to yield respective memory addresses. First, most expressions constructed by Econst obviously cannot be employed as the l-value because most of these represent a Lolisa constant value at the expression level directly. For brevity, we assign $\mathscr{A}]\sigma, *]$ to denote the recursive processes of array and map employed for searching the indexed addresses. Note that struct and field are forbidden to specify expressions in the l-value position to ensure that Lolisa is well-formed and well-behaved. The only means allowed in Lolisa of altering the fields of structures are using Estruct to either change all fields or declaring a new field. Although this limitation may be not friendly for programmers or verifiers, it avoids potential risks.

In the previous section, we defined the semantics of array values. Accordingly, we can define the address searching process based on the semantics of arrays as rule 9, which takes name, $\Downarrow_{\text{id}_{\text{array}}}$, $\tau$, and $\text{address}_{\text{offset}}$ as parameters. Similarly, we can define rule 10 below for mapping values:

$$\mathscr{A}]\sigma, \text{name}, \Downarrow_{id_{\text{array}}}, \tau, \text{address}_{\text{offset}}\Big], \tag{9}$$

$$\mathscr{A}]\sigma, \text{name}, \Downarrow_{id_{\text{map}}}, \tau, \tau_{\text{map}}, \text{snd}\Big]. \tag{10}$$

*5.1.2. Evaluating Expressions in the R-Value Position.* In the following, we assign $\Downarrow_{e}^{r}$ to denote the evaluation of expressions in the r-value position to yield the respective memory addresses.

As shown in Figure 10, the rules EVAL-REXP-CONS define the evaluation of constant expressions. Here, we note that, because constant expressions store Lolisa values directly, the results can be obtained by applying $\Downarrow_{\text{val}}$ directly. In the expression level, the r-value position is specified with a struct type. This is also the only means of initializing or changing the value of a struct-type term. The rules EVAL-REXP-STR defines this process. Here, if the evaluation of Estruct fails, the process of evaluating a member's value yields an error message. Otherwise, the member's value set is obtained and the respective struct memory value is returned. Finally, the semantics of binary and unary operations are defined according to the rules EVAL-REXP-BOP and EVAL-REXP-UOP.

Due to the static type limitations in the formal abstract syntax definition based on GADTs, the expressions, subexpressions, and operations are all guaranteed to be well-formed, and the type dependence relations need not be checked using, e.g., informal assistant functions, as required by other formal semantics such as Clight. The functions $\text{eval}_{\text{bop}}$ and $\text{eval}_{\text{uop}}$ take the results of expression evaluations and required operations as arguments, and combine them together to generate new memory values.

*5.2. Evaluation of Statements.* In the following, we assign $\Downarrow_{\text{stt}}$ to denote the evaluation process of statements, and parts of the necessary operational semantics are summarized in Figure 11. Most evaluations employ the helper functions

$$\frac{\begin{array}{c} M\vdash\sigma,b_{infor} \quad \mathcal{F},b_{infor}\vdash v \quad \mathcal{C}\vdash inhertis_c,con_{infor} \quad \Lambda\vdash inherits \\ \mathcal{E}\vdash env,fenv \quad \Lambda\vdash Econ(Some\ \alpha) \quad \mathcal{E},M,\mathcal{F}\vdash s \\ \mathcal{E},M,\mathcal{F}\vdash env_{check}(env,fenv)\hookrightarrow true \wedge inherit_{check}(inhertis_c,inherits)\hookrightarrow true \wedge set_{gas}(s,env)\hookrightarrow Some\ env' \\ write_{dir}\left(\left(\sigma,\alpha,Cid(\ (cid\ oaddr),infors,con_{infor},env,b_{infor})\right)\right)\hookrightarrow\sigma' \end{array}}{\mathcal{E},M,\mathcal{F}\vdash\langle\sigma,env,fenv,Contract\left(\left(Econ(Some\ \alpha)\right),inherits,s\right)\rangle\Rightarrow\langle\sigma',env',fenv,normal\rangle}\ (\text{EVAL-STT-CON})$$

$$\frac{\begin{array}{c} \mathcal{E}\vdash env,fenv \quad M\vdash\sigma,b_{infor} \quad \Lambda\vdash str_\tau \quad \Sigma,\Theta,\Lambda\vdash mems \\ \mathcal{E},M,\mathcal{F}\vdash env_{check}(env,fenv)\hookrightarrow true \wedge set_{gas}(s,env)\hookrightarrow Some\ env' \\ write_{dir}\left(\sigma,str_\tau,Str_{type}\left(str_\tau,mems,env,b_{infor}\right)\right)\hookrightarrow\sigma' \end{array}}{\mathcal{E},M,\mathcal{F}\vdash\ \langle\sigma,env,fenv,Struct\left(str_{type},mems\right)\rangle\Rightarrow\langle\sigma',env,fenv,stop\rangle}\ (\text{EVAL-STT- STRUCT})$$

$$\frac{\begin{array}{c} \mathcal{E}\vdash env,fenv \quad M\vdash\sigma,b_{infor} \quad \mathcal{E},M,\mathcal{F}\vdash e \quad \mathcal{E},M,\mathcal{F}\vdash inputs \\ \mathcal{E},M,\mathcal{F},b_{infor}\vdash Efun(Some\ \alpha) \\ env_{check}(env,fenv)\hookrightarrow true\ \wedge set_{gas}(s,env)\hookrightarrow Some\ env'\wedge read_{chck}\left(\sigma,env,b_{infor},\alpha\right)\hookrightarrow Some\ m_v \\ set_{env}(env,0,\alpha)\hookrightarrow env' \end{array}}{\mathcal{E},M,\mathcal{F}\vdash\langle\sigma,env,fenv,Fun_{call}\left(\left(Efun(Some\ \alpha)\right),inputs\right)\rangle\Rightarrow\langle\sigma\Downarrow_{m_v.(stt)},inputs,env',fenv,normal\rangle}\ (\text{EVAL-STT-FUN-CALL})$$

$$\frac{\begin{array}{c} \mathcal{E}\vdash env,fenv \quad \mathcal{E},M,\mathcal{F}\vdash fpars \quad M\vdash\sigma,b_{infor} \quad \mathcal{E},M,\mathcal{F}\vdash s \quad \mathcal{E},M,\mathcal{F}\vdash inputs \quad M\vdash\sigma \\ \mathcal{F},\Phi,b_{infor}\vdash Efun\left(Tundef,(Some\ a)\right) \\ env_{check}(env,fenv)\hookrightarrow true\wedge set_{gas}(Modifier(e,fpars,s),env)\hookrightarrow Some\ env' \\ \left(repeat\ init_{var}(\sigma,env,fenv,oacc,\tau,fpar)\right)\leftarrow fpars\equiv[fpar_0,fpar_1,\ldots,fpar_n]\hookrightarrow Some\ \sigma' \\ ①\ length(inputs)\leq length(fpars)\supset set_{par}\left(\sigma',fpars,inputs\right)\hookrightarrow Some\ \sigma'' \\ ②\ init_{re}(\sigma'',\Lambda_{fun},[Tundef])\hookrightarrow Some\ \sigma''' \\ ③\ write_{dir}(\ \sigma'',a,s)\hookrightarrow\sigma'''' \end{array}}{\mathcal{E},M,\mathcal{F},\Phi\vdash\langle\sigma,env,fenv,Modifier(e,fpars,s)\rangle\Rightarrow\langle\sigma''''\Downarrow_s,env',fenv,normal\rangle}\ (\text{EVAL-STT-MODI})$$

Figure 11: Part of formal statement semantics of Lolisa, including environment and gas checking, contract, struct, modifier, and function call statements.

$env_{check}$ and $set_{gas}$. The helper function $env_{check}$ takes the current environment env and the super-environment fenv as arguments, and checks conditions such as gas limitations and the congruence of execution levels. Contract declarations are one of the most important statements of Solidity. In Lolisa, contract declaration involves two operations. First, the consistency of inheritance information is checked using the helper function $inherit_{check}$, which takes the inheritance relations in module context $\mathcal{C}$ and the source code as arguments. Second, the initial contract information, including all member identifiers, is written into a designated memory block. As defined in Figure 11, the formal semantics of contract declaration are defined as EVAL-STT-CON below.

As rule EVAL-STT-STRUCT, the address is the new struct type identifier, and the struct-type information is written into the respective memory block directly.

In Lolisa, a function call statement is used to apply the function body indexed by the call statement. The process of applying an indexed function is defined by the rules EVAL-STT-FUN-CALL below.

Modifier declarations are a kind of special function declaration that requires three steps, and includes a single limitation. The parameter values are set by the $se_{tpar}$ predicate. As defined by the rule EVAL-STT-MODI in Figure 11, the first step (denoted as ①) initializes and sets the parameters. The second step (denoted as ②) stores the modifier body into the respective memory block. The third step (denoted as ③) attempts to initialize the return address $\Lambda_{fun}$. Due to the multiple return values, $init_{re}$ takes a return type list as an argument. Particularly, the modifier body can only yield an initial memory state, and therefore cannot change memory states. The difference between modifier semantics and function semantics is that function semantics include checking the modifier limitations restricting the function. Specifically, taking EVAL-STT-FUN as an example, before invoking a function, the modifier restricting the function will be executed. If the result of a modifier evaluation is $\sigma_{init}$, it means that the limitations checking of the modifier fails and the function invocation will be thrown out. Otherwise, the function will be executed.

*5.3. Development of Standard Library and Evaluation of Programs.* As discussed previously, we have developed a small standard library in Lolisa that incorporates the built-in data structures and functions of EVM to facilitate execution and verification of Solidity programs rewritten in Lolisa using higher-order logic theorem-proving assistants. Here, we discuss the standard library in detail. Then, based on the syntax, semantics, and standard library formalization, we define the semantics governing the evaluation (i.e., execution) of programs written in Lolisa.

*5.3.1. Development of the Standard Library and Evaluation of Programs.* Note that we assume the built-in data structures and functions of EVM are correct. This is reasonable because, first, the present focus is on verification of high-level smart contract applications rather than the correctness of EVM. Second, Lolisa is sufficiently powerful to implement any data structure or function employed by EVM. Thus, we only need to implement the logic of these built-in EVM features using Lolisa based on the Solidity documentation [4] to ensure that these features are well formed. For example, an address is a special compound type in Solidity that has the *balance*, *send*, and *call* members. However, we can treat an address as a special struct type in Lolisa and define it using the Lolisa syntax, as shown in Figure 12. All other built-in data structures and functions of EVM are defined in a similar manner. Typically, *requires* is a special standard

$$
\begin{aligned}
Address \overset{\text{def}}{=} Str_{type}\ _0xaddress\ (&str_{mem}\ TInt\ (Nvar\ addr) \\
&str_{mem}\ TInt\ (Nvar\ balance) \\
&str_{mem}\ (Tfid\ (Some\ _Send))\ (Nvar\ send) \\
&str_{mem}\ TInt\ (Nvar\ gas)\ (str_nil))))\ _0xaddress\ 2\ occupy)
\end{aligned}
$$

FIGURE 12: Address type declaration in Solidity and its equivalent as a special struct type in Lolisa syntax.

function that does not need a special address and, according to the Solidity documentation, is defined in Lolisa as rule 11:

$$
\text{requires} \underset{=}{\text{def}} \lambda\left(\forall\, \tau_0\colon\ \text{type},\ e\colon\ \exp r_{\tau_0\ \text{Tbool}}\right).\ (\text{If}\ e\ (\text{Snil})\ (\text{Throw})). \tag{11}
$$

Next, we pack these data structures and functions together as a standard library in Lolisa, which is executed prior to executing user programs. Thus, all built-in functions and data structures of EVM can be formalized in Lolisa, which allows the low-level behavior of EVM to be effectively simulated rather than building a formal EVM. Currently, this standard library is a small subset that only includes *msg*, *address*, *block*, *send*, *call*, and *requires*.

*5.3.2. Program Evaluation.* The semantics governing the execution of a Lolisa program (denoted as $P(\text{stt})$) is defined by rules 12 and 13, where $\infty$ refers to infinite execution and $T$ represents the set of termination conditions for finite execution.

$$
\frac{\begin{array}{l}
\varepsilon \vdash \text{env}, \text{fenv}\ M \vdash \sigma, b_{\text{inf or}} \mathscr{F} \vdash \text{opars}\ \varepsilon, M, \mathscr{F} \vdash P(\text{stt}) \varepsilon, M, \mathscr{F} \vdash \text{lib} \\
\text{env} = \text{set}_{\text{gas}}\left(\text{init}_{\text{env}}\left(P(\text{stt})\right)\right) \text{fenv} = \text{init}_{\text{env}}\left(P(\text{stt})\right) \\
\sigma = \text{init}_{\text{mem}}\left(P(\text{stt}), \text{lib}\right)
\end{array}}{\varepsilon, M, \mathscr{F} \vdash \langle \sigma, \text{env}, \text{fenv}, \text{opars}, \Downarrow_{P(\text{stt})} \rangle \Rightarrow \langle \sigma', \text{env}, \text{fenv} \rangle}, \tag{12}
$$

$$
\frac{\begin{array}{l}
\varepsilon \vdash \text{env}, \text{fenv}\ M \vdash \sigma, b_{\text{inf or}} \mathscr{F} \vdash \text{opars}\ \varepsilon, M, \mathscr{F} \vdash P(\text{stt}) \varepsilon, M, \mathscr{F} \vdash \text{lib} \\
\text{env} = \text{set}_{\text{gas}}\left(\text{init}_{\text{env}}\left(P(\text{stt})\right)\right) \text{fenv} = \text{init}_{\text{env}}\left(P(\text{stt})\right) \\
\sigma = \text{init}_{\text{mem}}\left(P(\text{stt}), \text{lib}\right)
\end{array}}{\varepsilon, M, \mathscr{F} \vdash \langle \sigma, \text{env}, \text{fenv}, \text{opars}, \Downarrow_{P(\text{stt})} \rangle \Rightarrow \langle \sigma', \text{env}', \text{fenv} \rangle \Lambda \text{env}'.(gas) \leq \text{fenv}.(\text{gasLimit}) \Rightarrow} \langle \sigma', \text{env}', \text{fenv} \rangle. \tag{13}
$$

These rules represent two conditions of $P(stt)$ execution. Under the first condition governed by rule 12, $P(\text{stt})$ terminates after a finite number of steps owing to a returned *stop*, *exit*, or *error*. Under the second condition governed by rule 13, $P(\text{stt})$ cannot terminate via its internal logic and would undergo an infinite number of steps. Therefore, $P(\text{stt})$ is deliberately stopped via the gas limitation checking mechanism. Here, *opars* represents a list of optional arguments. In addition, as discussed in Section 5.1, the initial environment env and super-environment fenv are equivalent, except for their gas values, which are initialized by the helper function $\text{init}_{\text{env}}$, and the initial gas value of env is set by $\text{set}_{\text{gas}}$. Finally, the initial memory state is set by $\text{init}_{\text{mem}}$, considering $P(\text{stt})$ and the standard library lib as arguments.

## 6. Formal Verification of Smart Contract Using FEther

As introduced in Section 1, we have implemented a formal verified interpreter in Coq for Lolisa, denoted as FEther [11], which incorporates about 7000 lines of Coq code (not including proofs and comments). This interpreter is developed strictly following the formal syntax and semantics of Lolisa based on the GERM framework. To be specific, FEther is implemented by computational functions (considered as the mechanized computational semantics), which are equivalent to the natural semantics of Lolisa given in this paper. The implementation is conducted following the details presented in our previous study [11] using Gallina, which is the functional programming language provided by Coq. Accordingly, FEther can parse the syntax of Lolisa to symbolically execute formal programs written in Lolisa. While efforts are ongoing to prove the consistency between the semantics of FEther and Lolisa, FEther can be employed to prove the properties of real-world programs. This process is effective at exposing errors not only in the test suites that exemplify expected behaviors but also in normal smart contracts. Specifically, a simple case study is presented to demonstrate the symbolic execution and verification process based on Lolisa and FEther. Its source code is presented in Appendix A, and the respective formal version written in Lolisa is presented in Appendix B. Here, it is clear that the program will be thrown out if the message sender in the index mapping list and the current time now are less than *privilegeOpen* or are greater than *privilegeClose*. This is easily proven manually with the inductive predicate semantics defined previously.

```
Lemma no_in_time : forall pump pump_val m m' m0 m1 m2 m3 z blc gs (env : environment) (s : statement),
  let (_, _, cur, dn) := env in
  m = init_msg init_m z 19 blc gs IcoController msg ->
  m' = write_by_address m (Map priviledges
          (Some (iStr _0xaddress (Some (cons (iInt (Some (INT I64 Signed 19)) public occupy)
                 (cons (iFid (fid (Some _0xsend)) (Some nil) public occupy)
                 (cons (iInt (Some (INT I64 Signed blc)) public occupy)
                 (cons (iInt (Some (INT I64 Signed gs)) public occupy)
                  nil))))) public occupy, Bool (Some true) cur 2 public occupy))
                  Iaddress Tbool None cur dn public occupy) priviledges ->
  m0 = write_by_address m' (Int (Some (INT I64 Unsigned 0)) cur 2 public occupy) privilegeOpen ->
  m1 = write_by_address m0 (Int (Some (INT I64 Unsigned 3)) cur 2 public occupy) privilegeClose ->
  m2 = write_by_address m1 (Int (Some (INT I64 Unsigned 4)) cur 2 public occupy) now ->
  m3 = mem_address (mem_msg m2) ->
  pump > 100 ->
  pump_val > 100 ->
  test pump pump_val m3 (Some nil) env env example
   = Some init_m'
.
Proof.
  destruct env0; intros. unfold mem_msg, mem_address in *. initmem.
  rewrite H4; clear H4.
  do 7 (step; unfold_modify_s).
  step.
  next pump_val.
  push. unfold init_addr_str; cbn in *.
  do 5 (next pump_val).
  push.
  do 4 (step; unfold_modify_s).
  repeat step.
Qed.
```

Figure 13: Execution and verification of the Lolisa program in Appendix B using the formal interpreter FEther in Coq.

Meanwhile, we can verify this property by symbolically executing the program with the help of FEther in Coq directly, as shown in Figure 13. The formal intermediate memory states obtained during the execution and verification of this Lolisa program using FEther are shown in Figure 14. Then, we can compare the mechanized verification results and the manually obtained results to validate the semantics of Lolisa. In addition, the application of FEther based on Lolisa and the GERM framework also certifies that our proposed EVI theory is feasible.

## 7. Discussion

*7.1. Contributions.* First, Lolisa formalizes most of the types, operators, and mechanisms of Solidity, and it includes most of the Solidity syntax. In addition, a standard library was built based on Lolisa to represent the built-in data structures and functions of EVM, such as msg, block, and send. As such, programs written in Solidity can be translated into Lolisa, and vice versa, with a line-by-line correspondence without rebuilding or abstracting, which are operations that can negatively impact consistency.

Second, the formal syntax in Lolisa is defined using generalized algebraic datatypes, which impart static type annotation to all the values and expressions of Lolisa. In this way, Lolisa has a stronger static type system than Solidity for checking the construction of programs. As such, it is impossible to construct ill-typed terms in Lolisa, which also assists in discovering ill-typed terms in Solidity source code. Moreover, the formal syntax ensures that all expressions and values in Lolisa are deterministic.

Finally, the syntax and semantics of Lolisa are mechanized using the Coq proof assistant. Besides, a formal verified interpreter FEther is developed in Coq to validate whether Lolisa satisfies the above Executable and *Provable* feature and the meta-properties of the semantics. In contrast to similar efforts focused on building formal syntax and semantics for high-level programming languages, the formal semantics of Lolisa are defined based on the FSPVM-E framework. As such, it is possible for programs written in Lolisa to be symbolically executed and have their properties simultaneously verified automatically in Coq proof assistant directly as program execution in the real world when conducted in conjunction with FEther.

*7.2. Limitations.* Although the novel features in the current version of Lolisa specification language confer a number of advantages, some limitations remain.

First, because the Lolisa is large subset of Solidity, some of Solidity characteristics, such as inline assembly, have been omitted in Lolisa. Hence, some complicated Ethereum smart contracts are not supported by the current version of Lolisa current. These characteristics will be supported in the updated version of Lolisa.

Second, the Lolisa is formalized at the Solidity source-code level. Although it will analyze vulnerabilities before the compiling process, it cannot guarantee the correctness of the corresponding bytecode when the compiler is untrusted. One possible solution is developing a low-level version of Lolisa, which executes the bytecode generated by the compiler, then proving the equivalence between Solidity execution results and the respective execution results of the bytecode.

Finally, although the current version of Lolisa can be verified in FEther symbolically, this process is not yet fully automated. In occasional situations, programmers must analyze the current proof goal and choose suitable verification tactics. Fortunately, this goal can be achieved by optimizing the design of the tactic evaluation strategies.

```
1 subgoal
pump, pump_val : nat
m, m', m0, m1, m2, m3 : memory
z, blc, gs : Z
l : list address
o : option address
a0 : address
d : dnum
s : statement
H5 : S pump > 100
H6 : pump_val > 100
______________________________________(1/1)
test pump pump_val
  {|
  m_init := initData;
  m_send := initData;
  m_send_re := initData;
  m_msg := Str_type _0xmsg
             (str_mem Taddress (Nvar sender)
                (str_mem Tint (Nvar values) str_nil)) _0xmsg 3 occupy;
  m_address := Str_type _0xaddress
                 (str_mem Tint (Nvar addr)
                    (str_mem Tint (Nvar balance)
                       (str_mem (Tfid (Some _Send)) (Nvar send)
                          (str_mem Tint (Nvar gas) str_nil)))) _0xaddress 3
                 occupy;
  m_throw := false;
  m_0x00000000 := initData;
  m_0x00000001 := initData;
  m_0x00000002 := initData;
  m_0x00000003 := initData;
  m 0x00000004 := initData;
  m_0x0000005f := initData;
  m_0x00000060 := initData;
  m_0x00000061 := initData;
  m_0x00000062 := initData;
  m_0x00000063 := initData |} (Some nil) (Evn l o a0 d) (Evn l o a0 d)
  (Var (Some public) (Evar (Some close) Tuint);;
   Var (Some public) (Evar (Some quota) Tuint);;
   Var (Some public) (Evar (Some rate) Tuint);;
   Var (Some public) (Evar (Some partiLimit) Tuint);;
   Var (Some public) (Evar (Some totalLimit) Tuint);;
   Var (Some public) (Evar (Some finalLimit) Tuint);;
   If
     (Econst (Vmap priviledges (Mstr_id Iaddress msg (sender ~>> \\\)) None))
     (Assignv (Evar (Some open) Tuint) (Evar (Some privilegeOpen) Tuint);;
      Assignv (Evar (Some close) Tuint) (Evar (Some privilegeClose) Tuint);;
      Assignv (Evar (Some quota) Tuint) (Evar (Some privilegeQuota) Tuint);;
      Assignv (Evar (Some rate) Tuint) RATE_PRIVILEGE;; nil)
     (Assignv (Evar (Some open) Tuint) (Evar (Some ordinaryOpen) Tuint);;
      Assignv (Evar (Some close) Tuint) (Evar (Some ordinaryClose) Tuint);;
      Assignv (Evar (Some quota) Tuint) (Evar (Some ordinaryQuota) Tuint);;
      Assignv (Evar (Some rate) Tuint) RATE_ORDINARY;; nil);;
   If
     (Evar (Some now) Tuint (<) Evar (Some open) Tuint
      (||) Evar (Some now) Tuint (>) Evar (Some close) Tuint) (Throw;; nil)
     (Snil;; nil);;
   If
     (Evar (Some subscription) Tuint (+) Evar (Some rate) Tuint
      (>) TOKEN_TARGET_AMOUNT) (Throw;; nil) (Snil;; nil);;
   If (Evar (Some index) Tuint (==) Econst (Vint (INT I64 Unsigned 0)))
     (Throw;; nil) (Snil;; nil);;
   If
     (Econst (Vmap deposits (Mvar_id Iuint index) None)
```

FIGURE 14: Formal memory states during the execution and verification of the Lolisa program in Appendix B using FEther in Coq.

## 8. Conclusion and Future Work

In this paper, we defined the formal syntax and semantics for a large subset of Solidity, which we denoted as Lolisa. The formal syntax of Lolisa is strongly typed according to GADTs. The syntax of Lolisa includes nearly all the syntax in Solidity, and the two languages are therefore equivalent with each other. As such, Solidity programs can be translated to Lolisa line-by-line without rebuilding or abstracting, which are operations that are too complex to be conducted by general programmers, and may introduce inconsistencies. Moreover, we have mechanized Lolisa in Coq completely, and have developed a formal interpreter FEther in mathematical tool Coq based on Lolisa, which was employed to validate the semantics of Lolisa. By basing the formal semantics of Lolisa on our FSPVM-E framework [31], programs written in Lolisa can be symbolically and automatically executed in Coq, and thereby verify the corresponding Solidity programs simultaneously. As a result of the present work, we can now directly verify smart contracts written in Solidity using Lolisa.

```
Solidity ˆ0.4.8;
function example () public payable {
    uint index = indexes[msg.sender];
    uint open;
    uint close; . . .
    if (privileges[msg.sender]) {
        open = privilegeOpen;
        close = privilegeClose;
    . . .} else {
        open = ordinaryOpen;
        close = ordinaryClose;. . .}
    if (now < open || now > close) {
        throw(); }
    if (subscription + rate > TOKEN_TARGET_AMOUNT) {
        throw (); }
    . . .
    if (msg.value <= finalLimit) {
        safe.transfer(msg.value);
        deposits[index] + = msg.value;
        subscription + = msg.value / 1000000000000000000 ∗ rate;
        Transfer(msg.sender, msg.value); } else {
        safe.transfer(finalLimit);
        deposits[index] + = finalLimit;
        subscription + = finalLimit / 1000000000000000000 ∗ rate;
        Transfer(msg.sender, finalLimit);
        msg.sender.transfer(msg.value - finalLimit);
    }
}
```

Algorithm 1: Partial source code of case study contract.

```
Coq ˆ8.8;
Definition Example : =
    (Fun public payable (Efun (Some example) Tundef) pnil nil);;
        (Var (Some public) Evar (Some index) Tuint));;
        (Assignv (Evar (Some index) Tuint)
        (Econst (@Vmap Iaddress Tuint indexes (Mstr_id Iaddress msg (sender ~>>\\\)) None));;
            (Var (Some public) (Evar (Some open) Tuint));;
            (Var (Some public) (Evar (Some close) Tuint));;
            (Var (Some public) (Evar (Some quota) Tuint));;
            . . .
            (If (Econst (@Vmap Iaddress Tbool priviledges
                (Mstr_id Iaddress msg (sender ~>>\\\)) None))
            ((Assignv (Evar (Some open) Tuint) (Evar (Some privilegeOpen) Tuint));;
                (Assignv (Evar (Some close) Tuint) (Evar (Some privilegeClose) Tuint));;
            . . .;; nil)
            ((Assignv (Evar (Some open) Tuint) (Evar (Some ordinaryOpen) Tuint));;
                (Assignv (Evar (Some close) Tuint) (Evar (Some ordinaryClose) Tuint));;
            . . .;; nil)
            (If ((Evar (Some now) Tuint) (<) (Evar (Some open) Tuint) (||)
            (Evar (Some now) Tuint) (>) (Evar (Some close) Tuint))
            (Throw;; nil) (Snil;; nil));;
            (If ((Evar (Some subscription) Tuint) (+) (Evar (Some rate) Tuint) (>)
                TOKEN_TARGET_AMOUNT)
            (Throw;; nil) (Snil;; nil));;
            . . .
            (If ((Econst (Vfield Tuint (Fstruct _0xmsg msg) (values ~> \\) None))
            (<=) (Evar (Some finalLimit) Tuint))
```

Algorithm 2: Continued.

```
((Fun_call (Econst (Vfield (Tfid (Some safe)) (Fstruct _0xaddress safe) (send ~> \\) None))
  (pccons (Econst (Vfield Tuint (Fstruct _0xmsg msg) (values ~> \\) None)) pcnil));;
    (Assignv (Econst (@Vmap Iuint Tuint deposits (Mvar_id Iuint index) None))
      ((Econst (Vfield Tuint (Fstruct _0xmsg msg) (values ~> \\) None)) (+)
      ((Econst (@Vmap Iuint Tuint deposits (Mar_id Iuint index) None))));;
  (Assignv (Evar (Some subscription) Tuint) ((Econst Vfield Tuint (Fstruct _0xmsg msg)
    (values ~> \\) None)) (+) (Evar (Some finalLimit) Tuint) (/)
    (Econst (Vint (INT I64 Unsigned 1000 000 000 000 000 000))) (x)
    (Evar (Some rate) Tuint))));; nil) . . .;; nil);; nil.
```

Algorithm 2: Formal version of Algorithm A written in Lolisa.

The source files containing the formalization of Lolisa abstract syntax tree are accessible at https://gitee.com/UESTC_EOS_FV/LolisaAST/tree/master/SPEC

Presently, we are working toward verifying the correctness of FEther, and developing a proof of the equivalence between computable semantics and inductive semantics. Subsequently, we will implement our proposed preliminary scheme based on the notation mechanism of Coq to extend Lolisa along two important avenues.

Our ongoing project is the extension of FSPVM-E to support EOS blockchain platform [32], and we will then verify our new framework in Coq. Besides, we will develop a general formal verification toolchain using HOL proof technology for blockchain smart contracts with the goal of automatic smart contract verification.

## Appendix

### A. Source Code of the Case Study

As shown in Algorithm 1, we give the partial source code of the case study contract.

### B. Formal Version of the Case Study

As shown in Algorithm 2, we give the formal version of Algorithm 1 written in Lolisa.

## Data Availability

The source files containing the formalization of Lolisa abstract syntax tree data used to support the findings of this study have been deposited in the Gitee repository (https://gitee.com/UESTC_EOS_FV/LolisaAST/tree/master/SPEC).

## Conflicts of Interest

The authors declare no conflict of interest.

## References

[1] S. B. Nakamoto, “A peer-to-peer electronic cash system,” 2020, https://bitcoin.org/bitcoin.pdf.

[2] A. Narayanan, J. Bonneau, E. Felten, A. Miller, and S. Goldfede, *Bitcoin and Cryptocurrency Technologies: A Comprehensive Introduction*, Princeton University Press, Princeton, NJ, USA, 1 edition, 2016.

[3] V. E. Buterin, “A next-generation smart contract and decentralized application platform,” 2020, https://github.com/ethereum/wiki/wiki/White-Paper.

[4] S. Demirkan, I. Demirkan, and A. McKee, “Blockchain technology in the future of business cyber security and accounting,” *Journal of Management Analytics*, vol. 7, no. 2, pp. 189–208, 2020.

[5] 2020 Ethereum solidity documentation. https://Solidity.readthedocs.io/en/develop/.

[6] J. McKee, J. Cheng, N. Xiong, L. Zhan, and Y. Zhang, “A distributed privacy preservation approach for big data in public health emergencies using smart contract and SGX,” *Computers Materials & Continua*, vol. 65, no. 1, pp. 723–741, 2020.

[7] 2020 The D. A. O. Attacked: Code issue leads to $60 million ether theft. https://www.coindesk. com/dao-attacked-code-issue-leads-60-million-ether-theft/.

[8] L. Luu, D. H. Chu, H. Olickel, P. Saxena, and A. Hobor, “Making smart contracts smarter,” in *Proceedings of ACM Conference on Computer and Communications Security*, pp. 24–28, Vienna, Austria, October 2016.

[9] H. Xi, C. Chen, and G. Chen, “Guarded recursive datatype constructors,” in *Proceedings of SIGPLAN-SIGACT Symposium on Principles of Programming Languages*, pp. 224–235, New Orleans, LA, USA, January 2003.

[10] 2020 The Coq proof assistant reference manual. https://coq.inria.fr/distrib/current/refman/.

[11] Z. Yang and H. Lei, “FEther: an extensible definitional interpreter for smart-contract verifications in Coq,” *IEEE Access*, vol. 7, pp. 37770–37791, 2019.

[12] T. M. Bits, https://github.com/trailofbits/manticore, 2020.

[13] B. M. Mueller, https://github.com/b-mueller/mythril/, 2020.

[14] E. Hildenbrandt, M. Saxena, X. Zhu, N. Rodrigues, and P. Daian, “KEVM: a complete semantics of the ethereum virtual machine,” in *Proceedings of IEEE Computer Security Foundations Symposium*, pp. 204–217, Oxford, UK, July 2018.

[15] Y. Hirai, “Defining the Ethereum Virtual Machine for interactive theorem provers,” in *Proceedings of Financial Cryptography and Data Security*, pp. 35–47, Sliema, Malta, April 2017.

[16] S. Amani, M. Bégel, and M. Bortin, “Towards verifying ethereum smart contract bytecode in Isabelle/HOL,” in *Proceedings of Acm Sigplan International Conference on Certified Programs*, pp. 66–77, Los Angeles, CA, USA, January 2018.

[17] C. Reitwiessner, “Dev update: formal methods,” 2020, https://ethereum.org/2016/09/01/formal-methods-roadmap/.

[18] P. Rizzo, “Ethereum seeks smart contract certainty,” 2020, http://www.coindesk.com/ethereum-formal-verification-smart-contracts/.

[19] C. Liu, H. Liu, Z. Cao, Z. Chen, B. Chen, and B. Roscoe, "Reguard: finding reentrancy bugs in smart contracts," in *Proceedings of IEEE/ACM International Conference on Software Engineering: Companion*, pp. 65–68, Gothenburg, Sweden, May 2018.

[20] P. Tsankov, A. M. Dan, D. Drachsler-Cohen, A. Gervais, F. Bünzli, and M. T. Vechev, "Securify: practical security analysis of smart contracts," in *Proceedings of ACM SIGSAC Conference on Computer and Communications Security*, pp. 67–82, Toronto, ON, Canada, October 2018.

[21] N. Grech, M. Kong, A. Jurisevic, L. Brent, B. Scholz, and Y. Smaragdakis, "Madmax: surviving out-of-gas conditions in ethereum smart contracts," in *Proceedings of the ACM on Programming Languages*, pp. 1–27, Philadelphia, PA, USA, November 2018.

[22] E. Albert, P. Gordillo, B. Livshits, A. Rubio, and I. E. Sergey, "A framework for high-level analysis of ethereum bytecode," in *Proceedings of Automated Technology for Verification and Analysis*, pp. 513–520, Los Angeles, CA, USA, October 2018.

[23] A. Mavridou, A. Laszka, E. Stachtiari, and A. Dubey, "Verisolid: correct-by-design smart contracts for ethereum," in *Proceedings of Financial Cryptography and Data Security*, pp. 446–465, Frigate Bay, St. Kitts and Nevis, February 2019.

[24] A. Mavridou and A. Laszka, "Tool demonstration: fsolidm for designing secure ethereum smart contracts," *Lecture Notes in Computer Science*, vol. 10804, pp. 270–277, Springer, Cham, Switzerland, 2018.

[25] T. Abdellatif and K. Brousmiche, "Formal verification of smart contracts based on users and blockchain behaviors models," in *Proceedings of International Conference on New Technologies, Mobility and Security*, pp. 1–5, Paris, France, February 2018.

[26] D. Park, A. Stefănescu, and G. Roşu, "KJS: a complete formal semantics of JavaScript," *Acm Sigplan Notices*, vol. 50, no. 6, pp. 346–356, 2015.

[27] X. Leroy, S. Blazy, D. Kästner, B. Schommer, and M. Pister, "Compcert-a formally verified optimizing compiler," in *Proceedings of European Congress on Embedded Real Time Software and Systems*, pp. 35–62, Toulouse, France, January 2016.

[28] A. W. Appel, *Verified Software Toolchain*, Springer Press, Berlin, Germany, 1 edition, 2011.

[29] R. H. Gu, Z. Shao, H. Chen et al., "CertiKOS: an extensible architecture for building cerified concurrent OS kernels," in *Proceedings of the USENIX Symposium on Operating Systems Design and Implementation*, pp. 653–669, Savannah, GA, USA, November 2016.

[30] S. Maffeis, J. C. Mitchell, and A. Taly, "An operational semantics for JavaScript," in *Proceedings of Asian Symposium on Programming Languages and Systems*, pp. 307–325, Bangalore, India, December 2008.

[31] Z. Yang, H. Lei, and W. Qian, "A hybrid formal verification system in Coq for ensuring the reliability and security of ethereum-based service smart contracts," *IEEE Access*, vol. 8, pp. 21411–21436, 2020.

[32] 2020 EOS blockchain documentation. https://eos.io/.